\documentclass[twocolumn,showpacs,amsmath,amssymb,aps,prd]{revtex4}

\usepackage{bm}

\begin{document}

\title{Quantization of the black hole area as quantization of the angular momentum component}

\author{Kostyantyn Ropotenko}
 \email{ro@stc.gov.ua}

\affiliation{State Administration of Communications, Ministry of
Transport and Communications of Ukraine,
\\ 22, Khreschatyk, 01001,
Kyiv, Ukraine}

\date{\today}

\begin{abstract}
In transforming from Schwarzschild to Euclidean Rindler coordinates
the Schwarzschild time transforms to a periodic angle. As is
well-known, this allows one to introduce the Hawking temperature and
is an origin of black hole thermodynamics. On the other hand,
according to quantum mechanics this angle is conjugate to the $z$
component of the angular momentum. From the commutation relation and
quantization condition for the angular momentum component it is
found that the area of the horizon of a Schwarzschild black hole is
quantized with the quantum $\Delta A = 8\pi l_P^{2}$. It is shown
that this conclusion is also valid for a generic Kerr-Newman black
hole.
\end{abstract}


\maketitle

 \section{Introduction}

Quantum gravity is understood to be a field of theoretical physics
attempting to unify quantum mechanics with general relativity in a
self-consistent manner. Unfortunately, we do not have yet a complete
and consistent quantum theory of gravity. It is widely believed
however that quantum gravity, as its title implies, should mean the
quantization of spacetime quantities. Obviously those effects of the
quantization should most strikingly manifest themselves in the
strong gravitational fields. Such conditions are provided in the
black holes. Bekenstein \cite{bek} was the first to suggest that
black holes should have a well-defined entropy associated with the
black hole area, and it was Bekenstein who first assumed that in
quantum gravity the black hole area should be quantized. By proving
that the black hole area is an adiabatic invariant, Bekenstein
showed that the area spectrum of black hole is of the form
\begin{equation}
\label{bek} A_n=\Delta A \cdot n,\quad n=0,1,2,...,
\end{equation}
where $\Delta A $ is the quantum of black hole area. Despite this
classical result there is still no general agreement on the precise
value of $\Delta A $; Bekenstein  suggested himself that $\Delta A =
8\pi l_P^{2}$ \cite{bek}.

In this paper I derive the quantization of black hole area
(\ref{bek}) and value of $\Delta A $ from the quantization of the
angular momentum component related to the Euclidean Rindler space of
a black hole. My approach does not depend on any particular theory
or model (e.g., string theory or loop quantum gravity). It is based
on the standard commutation rules of quantum mechanics for the
well-known black hole observables and utilizes the fundamental
properties of the Euclidean Rindler space underlying the black hole
thermodynamics.  It is well established \cite{wald,frol} that in the
near-horizon approximation the metric of an arbitrary black hole can
be reduced to the Rindler form. For these reasons, as will be
demonstrated below, my approach enables us to derive in a simple,
uniform way the black hole area spectrum for all kinds of black
holes and discover its universality. As a result, my conclusion is
valid not only for the Schwarzschild black hole, but also for all
other types of black holes and is universal.

The organization of this paper is as follows. In Sec. II we shall
introduce the Euclidean Rindler space, define the angular momentum
component, find its eigenvalues, and derive the area spectrum for a
Schwarzschild black hole. Finally, in Sec. III, following Sec. II,
we shall derive the area spectrum for other types of black holes and
demonstrate its universality.

\section{Quantization of area for
a Schwarzschild black hole}
\subsection{Euclidean Rindler
space}

For this purpose let us first consider a nonrotating uncharged black
hole with the mass $M$ and the Schwarzschild metric
\begin{equation}
\label{metr1}ds^{2}=-\left(1-\frac{2GM}{r}\right)dt^{2}+
\left(1-\frac{2GM}{r}\right)^{-1}dr^{2}+r^{2}d\Omega^{2},
\end{equation}
where all quantities have the standard meaning. In classical gravity
there exist two very important quantities related to the metric -
the black hole energy (mass) $M$ and Schwarzschild time $t$. It is
generally accepted that the black hole energy (mass) is a
well-defined observable. We adopt this assumption. In addition we
shall assume that the Schwarzschild time $t$ is also an observable.
The validity of these natural assumptions will not be debated here
(see \cite{olkh} and references therein). Then, according to the
quantum-mechanical correspondence principle we can replace $M$ and
$t$ by the quantum-mechanical operators of energy (associated with
the Killing operator of time translations)
\begin{equation}
  \hat{M} \equiv
  \left\{
  \begin{array}{cll}
    i\hbar \partial /\partial t, & \mbox{in the time ($t$)- representation}, \\
    M, & \mbox{in the energy representation,}
  \end{array}
  \right.
\label{mom1}
\end{equation}
and time
\begin{equation}
  \hat{t} \equiv
  \left\{
  \begin{array}{cll}
    t, & \mbox{in the time ($t$)- representation}, \\
    -i\hbar\partial /\partial M, & \mbox{in the energy representation,}
  \end{array}
  \right.
\label{om2}
\end{equation}
respectively. We shall further imply that our operators act in a
Hilbert space of black hole states but we do not aim to develop a
complete theory of operators in this space: for our purpose we need
only the commutation rules for the relevant observables. In
particular, it is evident that $\hat{M}$ and $t$ are self-conjugate
and obey the standard commutation rule for energy and time
\begin{equation}
\label{com1}[\hat{M},t]=i\hbar.
\end{equation}

Let us now continue $t$ to imaginary values $t\rightarrow it$ to
convert the Schwarzschild metric (\ref{metr1}) to the Euclidean
Schwarzschild metric
\begin{equation}
\label{metr2}ds^{2}=+\left(1-\frac{2GM}{r}\right)dt^{2}+
\left(1-\frac{2GM}{r}\right)^{-1}dr^{2}+r^{2}d\Omega^{2}.
\end{equation}
Introducing the coordinate $x$ defined by $r = 2GM + x^{2}/8GM$ and
expanding the Schwarzschild metric around $2GM$ we obtain
\cite{wald,frol}
\begin{equation}
\label{metr3}ds_E^{2}\approx(kx)^{2}dt^{2}+dx^{2}+\frac{1}{4k^{2}}d\Omega^{2},
\end{equation}
where the constant $k$ coincides with the surface gravity of a
Schwarzschild black hole, $k = 1/4GM$. This metric is the product of
the metric on a two-sphere with radius $2GM$ (the last term) and the
Euclidean Rindler metric
\begin{equation}
\label{metr4}ds_E^{2}=x^{2}d(kt)^{2}+dx^{2}.
\end{equation}

\subsection{Quantization of the angular momentum component and the area spectrum}

The metric (\ref{metr4}) has a coordinate singularity at $x = 0$
(corresponding to $r = 2GM$). Regularity is obtained if $kt$ is
interpreted as an angular coordinate $\omega$ with periodicity
$2\pi$
\begin{equation}
\label{om} \omega=kt=\frac{t}{4GM}
\end{equation}
($t$ itself has then periodicity $8\pi GM$ which, when set equal to
$\hbar/T_H$, gives the Hawking temperature $T_H$). The angular
character of $\omega$ is crucial for our consideration; the great
importance of the Euclidean Rindler metric is just motivated by this
fact because in quantum mechanics for the angle $\omega$ there
exists a quantized conjugate observable  - the $z$ component of the
angular momentum. On the contrary, in the Minkowskian Rindler
metric, the coordinate $\omega$ is not periodic and has a timelike
character; for this reason it is called Rindler time and its
conjugate is the Rindler energy $\hat{E}_R\equiv
i\hbar\partial/\partial\omega$ \cite{sus}. Thus we can define the
two canonically conjugate operators - the operator of the angular
momentum component (we shall, for the sake of convenience, call it
Rindler angular momentum and denote by $\hat{R}_z$)
\begin{equation}
  \hat{R_z} \equiv
  \left\{
  \begin{array}{cll}
    i\hbar \partial /\partial \omega, & \mbox{in SPC representation}, \\
    R_z, & \mbox{in the momentum representation},
  \end{array}
  \right.
\label{mom1}
\end{equation}
and the operator
\begin{equation}
  \hat{\omega} \equiv
  \left\{
  \begin{array}{cll}
    \omega, & \mbox{in SPC representation}, \\
    -i\hbar\partial /\partial R_z, & \mbox{in the momentum
    representation},
  \end{array}
  \right.
\label{om2}
\end{equation}
so that
\begin{equation}
\label{com2}[\hat{R_z},\omega]=i\hbar.
\end{equation}
Here the abbreviation SPC stands for spherical polar coordinate. Now
let us rewrite (\ref{com2}) as
\begin{equation}
\label{com3}[\hat{R_z},\frac{t}{4GM}]=i\hbar
\end{equation}
or
\begin{equation}
\label{com4}[\hat{R_z},t]=i\hbar4GM.
\end{equation}
Then, taking into account the conjugate character of $M$ and $t$
(\ref{com1}), and that $\hat{t}\equiv-i\hbar
\partial/\partial M$ and $\hat{R_z} \equiv R_z$ in the energy
representation, we obtain from (\ref{com4})
\begin{equation}
\label{eq1} \frac{\partial R_z}{\partial M}=4GM.
\end{equation}
As is easily seen, this is nothing but the first law of black hole
thermodynamics
\begin{equation}
\label{eq2}\frac{\partial S}{\partial M}=\frac{1}{T_H}.
\end{equation}
Solving the differential equation (\ref{eq1}) with the natural
boundary condition
\begin{equation}
\label{bound1}R_z=0 \quad \mbox{at}\quad M=0
\end{equation}
we find
\begin{equation}
\label{sol1} R_z=2GM^{2}
\end{equation}
or
\begin{equation}
\label{sol2} R_z=\frac{A}{8\pi G},
\end{equation}
where $A$ is the black hole area related with the Schwarzschild
radius $R_g=2GM$ in the usual way, $A=4\pi R_g^{2}$.  According to
the rule of quantization of the angular momentum component in
quantum mechanics, the eigenvalues of $\hat{R}_z$ are the positive
and negative integers, including zero,  multiple of $\hbar$. The
negative integers correspond to the region $r<R_g$. Since the
Euclidean Rindler line element does not penetrate into the horizon,
the negative integers can be ruled out. So we have
\begin{equation}
\label{quant1} \frac{A}{8\pi G}=n\cdot \hbar,\quad n=0,1,2,...
\end{equation}
or
\begin{equation}
\label{quant2} \frac{A}{8\pi l_p^{2}}=n,\quad n=0,1,2,...\,.
\end{equation}
But this is nothing else than the quantization rule for the black
hole area (\ref{bek}). Thus we can conclude that the black hole area
of a Schwarzschild black hole is quantized and the area spectrum is
equidistant with the quantum $\Delta A = 8\pi l_P^{2}$. Note that
this value already has been proposed in the literature (see
\cite{med1} and references therein) but its connection with
quantization of the Rindler angular momentum has not. Accordingly,
the entropy spectrum is given by
\begin{equation}
\label{entr1} S_n=2\pi \cdot n,\quad n=0,1,2,...\,.
\end{equation}
This agrees with the old Bekenstein conjecture \cite{bek}. This also
agrees with the result obtained by Barvinsky and Kunstatter
\cite{bar1} in generic $2-D$ dilaton gravity. Remarkably, they used
Euclidean black holes and quantization of angle-action variables of
the theory. But the authors do not deal with the angle-action
variables directly. Instead of this they use variables that are
derived from the angle-action ones via some canonical transformation
with the help of the first law of black hole thermodynamics. As a
result, the authors obtain a Hamiltonian for a simple harmonic
oscillator expressed in terms of the "position" and "momentum"
operators and reduce the entropy spectrum problem to the eigenvalue
problem of the Hamiltonian. This really gives an equidistant area
spectrum but leads to the existence of the stable black hole
remnants (because of a zero-point value). Following the previous
work, Barvinsky, Das, and Kunstatter also derived the discrete
two-parameter area spectrum for a charged black hole \cite{bar2};
Gour and Medved \cite{gour} did the same for a rotating one.
Maggiore \cite{mag} found the equidistant area spectrum with the
quantum $\Delta A = 8\pi l_P^{2}$ for a Schwarzschild black hole
with apparently very different arguments. Maggiore used a refined
semiclassical Hod's approach \cite{hod}; Hod derived the equally
spaced area spectrum with the quantum $\Delta A = 4l_P^{2}\ln3$
using the Bohr correspondence principle and the complex spectrum of
the quasinormal modes that correspond to the perturbation equation
for a Schwarzschild black hole. Note that Kunstatter \cite{kun} also
obtained the same result as was found by Hod using the
Bohr-Sommerfeld quantization condition. Motivated by the work of
Maggiore, Vagenas \cite{vag} and also Medved \cite{med2} obtained
for a rotating Kerr black hole the same result as was found by
Maggiore for a Schwarzschild black hole. On these grounds the
authors concluded that their result is universal. It is interesting
that, in doing so, they used different (although semiclassical)
approaches of Hod and Kunstatter. As will be demonstrated below, our
approach provides a support for their result. But our approach is
completely quantum-mechanical. Moreover it does not depend on any
particular theory ($2-D$ dilaton gravity) and does not yield the
remnants. It only utilizes the fundamental properties of the
Euclidean Rindler space. For these reasons our conclusion is valid
not only for the Schwarzschild black hole, but also for all other
types of black holes, and is universal. Note that the case of the
extremal and near-extremal black holes will not be considered here.

\section{Quantization of area for
other types of black holes}

For this purpose consider an arbitrary four-dimensional black hole
with the mass $M$,  the charge $Q$, the angular moment $J$, and the
specific angular momentum $a=J/M$. It is well established
\cite{wald,frol} that in the near-horizon approximation the metric
of an arbitrary black hole can be reduced to the Rindler form. So we
can simply write the general two-dimensional part of the Euclidean
Rindler metric for an arbitrary black hole as
\begin{equation}
\label{metr5}ds_E^{2}=x^{2}d(kt)^{2}+dx^{2},
\end{equation}
where $k$ is the surface gravity, (in units $G=c=1$ for short)
\begin{equation}
\label{surgrav2} k=\frac{4\pi \sqrt{M^{2}-Q^{2}-a^{2}}}{A},
\end{equation}
and $A$ is the area of the black hole
\begin{equation}
\label{areaBH} A=4\pi (2M^{2}-Q^{2}+2M\sqrt{M^{2}-Q^{2}-a^{2}}).
\end{equation}
The metric (\ref{metr5}) has a conical coordinate singularity which
disappears if $kt$ has a period equal to $2\pi$. Then, the request
that the metric is smooth yields the Hawking temperature
$T_H=k/2\pi$. Now, as in the Schwarzschild case, we can introduce
the Rindler angular momentum $\hat{R}_z$ with the commutation
relation
\begin{equation}
\label{com5}[\hat{R_z},t]=i\hbar\frac{1}{k}
\end{equation}
and obtain the equation
\begin{equation}
\label{eq3} \frac{\partial R_z}{\partial M}=\frac{1}{k}.
\end{equation}
Solving the differential equation with the same boundary condition
we find
\begin{equation}
R_z \equiv \left\{
\begin{array}{cll}
2M^{2},& \mbox{for S bh},\\
    M^{2}+M\sqrt{M^{2}-a^{2}}, & \mbox{for K bh}, \\
    M^{2}-Q^{2}/2 +M\sqrt{M^{2}-Q^{2}}, & \mbox{for RN bh}, \\
    M^{2}-Q^{2}/2+M\sqrt{M^{2}-Q^{2}-a^{2}}, & \mbox{for KN bh},
  \end{array}
  \right.
\label{sol2}
\end{equation}
or
\begin{equation}
\label{sol11} R_z=\frac{A}{8\pi}.
\end{equation}
Here the abbreviations S, K, RN, KN stand for Schwarzschild, Kerr,
Reissner-Nordstr\"{o}m, and  Kerr-Newman, respectively, bh is black
hole. Then, according to quantum mechanics we can write (restoring
$G$)
\begin{equation}
\label{quant1} \frac{A}{8\pi G}=n\cdot \hbar,\quad n=0,1,2,...
\end{equation}
or
\begin{equation}
\label{quant2} \frac{A}{8\pi l_p^{2}}=n,\quad n=0,1,2,...\,.
\end{equation}
Therefore the black hole area of an arbitrary black hole is
quantized and the area spectrum is equally spaced with the quantum
$\Delta A = 8\pi l_P^{2}$. Accordingly, the entropy spectrum is
given by
\begin{equation}
\label{entr1} S_n=2\pi \cdot n,\quad n=0,1,2,...\,.
\end{equation}
Thus the quantization rule is really universal. From the
thermodynamical point of view, the universality lies in the first
law of black hole thermodynamics
\begin{equation}
\label{eq3}\frac{\partial R_z}{\partial M}=\frac{1}{k}
\leftrightarrow \frac{\partial S}{\partial M}=\frac{1}{T_H}.
\end{equation}
Note that the universality of the Kerr area spectrum was found by
Vagenas \cite{vag} and also Medved \cite{med2} just from an
"adiabatically invariant" part of the first law of black hole
thermodynamics. Of course, in general we should take into account
the quantization of the charge and the angular momentum (spin) of a
black hole. Then the generalized angular momentum will be made up of
the intrinsic Rindler angular momentum, the angular momentum of the
electromagnetic field of a black hole, and the spin.

\end{document}